\documentclass[useAMS,usenatbib]{mn2e}
\usepackage{rotating}

 \usepackage{times}

\def\pcm3{{\rm\thinspace cm^{-3}}}

\def\contcaption{\@conttrue\SFB@caption\@captype}

\def\n_h{{\rm n_{H}}}

\def\NH1{{$N_{\rm HI}~$}}

\def\ga{{\rm\thinspace gauss}}

\def\approxlt{\mathrel{\hbox{\rlap{\lower .5ex \hbox {$\sim$}}
        \raise .15 ex \hbox{$<$}}}}
\def\approxgt{\mathrel{\hbox{\rlap{\lower .5ex \hbox {$\sim$}}
        \raise .15 ex \hbox{$>$}}}}

\def\la{\mathrel{\hbox{\rlap{\hbox{\lower4pt\hbox{$\sim$}}}\hbox{$<$}}}}
\def\ga{\mathrel{\hbox{\rlap{\hbox{\lower4pt\hbox{$\sim$}}}\hbox{$>$}}}}

\newbox\grsign \setbox\grsign=\hbox{$>$} \newdimen\grdimen
\grdimen=\ht\grsign
\newbox\simlessbox \newbox\simgreatbox \newbox\simpropbox
\setbox\simgreatbox=\hbox{\raise.5ex\hbox{$>$}\llap
     {\lower.5ex\hbox{$\sim$}}}\ht1=\grdimen\dp1=0pt
\setbox\simlessbox=\hbox{\raise.5ex\hbox{$<$}\llap
     {\lower.5ex\hbox{$\sim$}}}\ht2=\grdimen\dp2=0pt
\setbox\simpropbox=\hbox{\raise.5ex\hbox{$\propto$}\llap
     {\lower.5ex\hbox{$\sim$}}}\ht2=\grdimen\dp2=0pt
\def\simgreat{\mathrel{\copy\simgreatbox}}
\def\simless{\mathrel{\copy\simlessbox}}

\title[The ultramassive white dwarf GD50]{On the origin of the ultramassive white dwarf GD50
$\thanks{Based on observations collected at the European
Southern Observatory, Chile. ESO No. 072.D-0362}$}

\author[P. D. Dobbie et al.]{P. D. Dobbie$^{1}$\thanks{E-mail:
pdd@star.le.ac.uk} R. Napiwotzki$^{2}$ N. Lodieu$^{1}$ M. R. Burleigh$^{1}$ M. A. Barstow$^{1}$ R. F. Jameson$^{1}$\\
$^{1}$Department of Physics and Astronomy, University of Leicester, University Road, Leicester LE1 7RH, UK\\
$^{2}$Centre for Astrophysics Research, Science \& Technology Research Institute, University of Hertfordshire,
College Lane, Hatfield, AL10 9AB \\}

\begin{document}

\date{Accepted 1988 December 15. Received 1988 December 14; in original form 1988 October 11}

\pagerange{\pageref{firstpage}--\pageref{lastpage}} \pubyear{2002}

\maketitle

\label{firstpage}

\begin{abstract}

We argue on the basis of astrometric and spectroscopic data that the ultramassive white dwarf 
GD50 is associated with the star formation event that created the Pleiades and is 
possibly a former member of this cluster. Its cooling age ($\sim$60Myrs) is consistent with 
it having evolved essentially as a single star from a progenitor with a mass M$>$6M$
_{\odot}$ so we find no need to invoke a white dwarf-white dwarf binary merger scenario to 
account for its existence. This result may represent the first direct observational evidence 
that single star evolution can produce white dwarfs with M$>$1.1M$_{\odot}$, as predicted by 
some stellar evolutionary theories. On the basis of its tangential velocity we also provisionally 
identify the ultramassive (M$\sim$1.2M$_{\odot}$) white dwarf PG0136+251 as being related to the Pleiades. 
These findings may help to alleviate the difficulties in reconciling the observed number of hot nearby 
ultramassive white dwarfs with the smaller number predicted by binary evolution models under the assumption
that they are the products of white dwarf mergers.

\end{abstract}

\begin{keywords}

stars: white dwarfs

\end{keywords}

\section{Introduction}

GD50 (WD0346-011) is a well studied nearby hot H-rich white dwarf with a mass of 
M$>$1.1M$_{\odot}$ (e.g. Giclas, Burnham \& Thomas 1965, Bergeron et al. 
1991; Table 1). Extrapolation of recent determinations of the initial mass-final 
mass relation (IFMR; e.g. Dobbie et al. 2006) suggest that if it evolved via 
single star evolution then its progenitor likely had a mass M$\simgreat$6M$_{\odot}$. However, 
it appears to reside well away from any known star forming region, stellar 
association or young ($\tau$$\simless$200Myrs) open star cluster.
This is somewhat surprising as current observational evidence indicates that
massive stars are born predominantly within rich stellar associations or 
star clusters (e.g. Lada \& Lada 2003, de Wit et al. 2004). 

More generally, the observed number of massive white dwarfs, M$>$0.8M$_{\odot}$, may be 
too large for them all to have evolved as single stars from OB type progenitors, for reasonable 
assumptions about the Galactic star formation history and the form of the IFMR. Accordingly,
it has been suggested
that GD50 and a large proportion of the massive white dwarfs may have formed 
through binary evolution ($\approx$80\%; Liebert et al. 2005a). For example,
these objects may be born from the merger of two white dwarfs (either He+CO 
or CO+CO) each with a mass closer to the canonical value of 0.6M$_{\odot}$ 
(e.g. Segretain, Chabrier \& Mochkovitch 1997, Mochkovitch 1993). The detection
in an Extreme Ultraviolet Explorer spectrum of GD50 of what appeared to be 
rotationally  broadened (1000kms$^{-1}$) absorption lines of photospheric 
helium lent credance to this hypothesis (Vennes et al. 1996) since it might 
be expected that these merger products are rapid rotators with unusual 
atmospheric compositions. However, a more recent and detailed high resolution
study of the spectral energy distribution of this star centered on the 
H-$\alpha$ line failed to corroborate this result and has set an upper limit 
on the rotational velocity of $v \sin i$$\simless$35kms$^{-1}$ (Vennes 1999). 
Furthermore, an extreme-ultraviolet spectroscopic survey of a further 
eight ultramassive white dwarfs failed to reveal compelling evidence for the 
presence of helium in the atmospheres of any of these objects (Dupuis, Vennes
\& Chayer 2002). Thus the origin of GD50 and ultramassive white dwarfs in 
general is still an open question.

In this short communication we argue, based on astrometric and spectroscopic 
data and our recent re-evaluation of the form of the IFMR, that GD50 is associated with 
the Pleiades open cluster and is possibly a former member with properties 
consistent with having evolved essentially as a single star. We briefly discuss 
the implications of this result for stellar evolution theory.
\begin{table*}
\begin{minipage}{170mm}
\begin{center}
\caption{Summary details of GD50 (WD0346$-$011) including coordinates, apparent visual magnitude (Marsh et 
al. 1997), distance, proper motion, radial velocity and heliocentric space velocity.}
\label{sum1}
\begin{tabular}{llcccccccc}
\hline
\multicolumn{1}{|c|}{RA}    &  \multicolumn{1}{|c|}{Dec}   & V & d &  $\mu_{\alpha}\cos\delta$ & $\mu_{\delta}$ & RV & U & V & W \\
\multicolumn{2}{|c|}{J2000.0} & & pc &\multicolumn{2}{|c|}{mas yr$^{-1}$} & kms$^{-1}$ & \multicolumn{3}{|c|}{kms$^{-1}$} \\
 \hline
03 48 50.20  & -00 58 31.2 & 14.04$\pm$0.02  & 31.0$\pm$1.7 & +86.6$\pm$6.8 & -164.4$\pm$5.0 & +13.8$\pm$12.2 &-3.8$\pm$9.2 & -28.0$\pm$2.2 & -11.8$\pm$7.9 \\% & +75.1$\pm$16.4 & -155.9$\pm$18.1
\hline
\end{tabular}
\end{center}
\end{minipage}
\end{table*}

\section[]{The heliocentric space velocity of GD50}

We have obtained estimates of the proper motion of GD50 from the SuperCOSMOS Sky Survey database 
($\mu_{\alpha}\cos\delta$=+75.1$\pm$16.4mas~yr$^{-1}$, $\mu_{\delta}$=$-$155.9$\pm$18.1mas~yr$^{-1}$; 
Hambly et al. 2001), the Lick Northern Proper Motion Program ($\mu_{\alpha}\cos\delta$=+89.4$\pm
$5.5mas~yr$^{-1}$, $\mu_{\delta}$=$-$165.6.9$\pm$5.5mas~yr$^{-1}$; Klemola et al. 1987) and from our 
own measurements of the Palomar Sky Sky (E932, 1953/12/31) and UK Schmidt Telescope (9597, 1986/09/22) 
survey plates ($\mu_{\alpha}\cos\delta$$\approx$+86$\pm$4mas~yr$^{-1}$, $\mu_{\delta}$$\approx$$-$164$
\pm6$mas~yr$^{-1}$). These values are consistent with each other within their respective uncertainties: 
we determine a weighted mean proper motion for GD50 of $\mu_{\alpha}\cos\delta$=+86.6$\pm$6.8mas~yr$^{-1}
$, $\mu_{\delta}$=$-$164.4$\pm$5.0mas~yr$^{-1}$, where the final uncertainties have been estimated from 
the scatter of the individual measurements.

From a high S/N UVES spectrum of the white dwarf taken in the course of the SPY programme (Napiwotzki
et al. 2003) we have determined the weighted mean redshift of the H-$\alpha$ and H-$\beta$ line cores
to be +176.0$\pm$4.3kms$^{-1}$, in excellent agreement with the earlier measurement of
Vennes (1999). Additionally, by determining the weighted means of the robust estimates
found in the refereed literature of the last 15 years and new measurements obtained 
from our independent analysis of the FORS1 spectrum of GD50 detailed in Aznar-Cuadrado
et al. (2004) we have estimated the effective temperature and surface gravity of this 
object to be T$_{\rm eff}$=41550$\pm$720K and log g=9.15$\pm$0.05, respectively. As 
listed in Table 2, the errors in the individual determinations of effective temperature 
and surface gravity have been assumed to be at least 2.3\% and 0.07~dex, respectively (e.g. see 
Napiwotzki et al. 1999), while the scatter of these measurements has been used to estimate the 
final uncertainties in the mean parameters. The measurements of Barstow et al. (1993) were 
rejected from the calculation as they were found to lie $>$3$\sigma$ in both 
effective temperature and surface gravity from an initial estimate of the means.

Using CO core, thick (thin) H-layer, white dwarf evolutionary models (e.g. Fontaine, Brassard 
\& Bergeron 2001) we infer the mass, the radius and the cooling time of GD50 to be 1.264$\pm
$0.017 (1.253$\pm$0.018)M$_{\odot}$, 0.00495$\pm$0.00034 (0.00493$\pm$0.00034)R$_{\odot}$ and
 61$\pm$6 (58$\pm$6)Myrs, respectively. We note that the specific choice of core composition, CO 
or ONe, is believed to make little difference these values (e.g. van Kerkwijk \& Kulkarni 1999, 
Hamada \& Salpeter 1961). We determine the gravitational contribution to the line 
core redshift to be +162.2$\pm$11.4 (+161.4$\pm$11.4)kms$^{-1}$. Thus we estimate the 
radial velocity of GD50 to be +13.8$\pm$12.2 (+14.6$\pm$12.2)kms$^{-1}$. Based on a 
modern grid of white dwarf synthetic photometry (Holberg \& Bergeron 2006; Bergeron et al. 1995),
 the above effective temperature and surface gravity estimates and the V magnitude 
from Marsh et al. (1997) we determine the distance of GD50 to be 31.0$\pm$1.7 (30.9$\pm$1.7)pc. 
Following the prescription given by Johnson \& Soderblom (1987) we calculate the heliocentric
space velocity of GD50 to be U=$-$3.8$\pm$9.2 ($-$4.4$\pm$9.2)kms$^{-1}$, V=$-$28.0$\pm$2.2 ($
-$28.1$\pm$2.2)kms$^{-1}$ and W=$-$11.8$\pm$7.9 ($-$12.3$\pm$7.9)kms$^{-1}$ (see Table 1). 

The recent kinematical study performed by Pauli et al. (2006) of 398 field white dwarfs drawn 
from the catalogue of McCook \& Sion (2000) reveals that $\sim$75\% have V$>$-25kms$^{-1}$. 
Additionally, the error bounds on the U,V and W velocity components of only 3 of these stars overlap with the 
range defined by the above values and their associated uncertainties (assuming a solar velocity with respect to the local standard of rest of U=10.00kms$^{-1}$, V=5.25kms$^{-1}$ and W=7.17kms$^{-1}$; Dehnen \& Binney 1998). Thus we conclude that GD50 lies 
in a region of velocity space which is not heavily populated by white dwarfs. However, we note that
the space velocity of this object is strikingly close to that of Pleiades open star cluster and 
also the AB Dor moving group (Zuckermann et al. 2004). The 
Pleiades is the closest rich young ($\tau$$\approx$125Myrs) open cluster to GD50. We determine
its space motion to be U=$-$6.5$\pm$0.5kms$^{-1}$, V=$-$27.8$\pm$1.1kms$^{-1}$ and W=$-$14.6$\pm
$0.5kms$^{-1}$, for a distance of 134$\pm$5pc (the weighted mean of recent ground based 
estimates e.g. Percival et al. 2005, Southworth et al. 2005, Pan et al. 2004, Munari et al. 2004, 
Zwahlen et al. 2004, Gatewood et al. 2000), a radial velocity of +5.7$\pm$0.5kms$^{-1}$ and a proper motion of 
$\mu_{\alpha}\cos\delta$=+19.15.1$\pm$0.23mas yr$^{-1}$,$\mu_{\delta}$=-45.72$\pm$0.18mas yr
$^{-1}$ (Robichon et al. 1999). Luhman et al. (2005) estimate the space motion of the AB 
Dor moving group, within which the Sun resides, to be U=$-$7.7$\pm$0.4kms$^{-1}$, V=$-$26.0$\pm$0.4kms
$^{-1}$ and W=$-$13.6$\pm$0.3kms$^{-1}$.

\section{Discussion}
\subsection{The Local Association} 

The Pleiades and the AB Dor moving group are part of the Local Association, a large scale 
``supercluster'' structure consisting of a gravitationally unbound collection of open clusters, 
stellar associations and moving groups with comparable space motions (e.g. Sco-Cen, IC2602, 
Per OB3, Cas-Tau and NGC2516; Eggen 1992). While this could be taken as evidence of a common 
ancestry (e.g. Weidemann et al. 1992), the age of the members of this supercluster is estimated
to span the relatively broad range of $\sim$few Myrs to $\sim$few 100Myrs (e.g. Eggen 1992, 
Asiain et al. 1999). Chereul et al. (1998) argue that the Local Association
may just be the result of the chance superposition in phase space of several smaller stellar 
streams, each associated with a star formation event which if sufficiently massive manufactured 
a gravitationally bound star cluster. Indeed, Hipparcos observations of
open clusters within the Local Association reveal that at a level of $<$10kms$^{-1}$ their space 
velocities are quite distinct e.g.  Pleiades, U=$-$6.5$\pm$0.5kms$^{-1}$, V=$-$27.8$\pm$1.1kms$
^{-1}$ and W=$-$14.6$\pm$0.5kms$^{-1}$, Per OB3 ($\alpha$-Per), U=$-$15.3$\pm$0.7kms$^{-1}$, V=$
-$25.8$\pm$1.0kms$^{-1}$, W=$-$7.9$\pm$0.4kms$^{-1}$) and NGC2516, U=$-$17.4$\pm$1.4kms$^{-1}$, V=$
-$23.7$\pm$0.4kms$^{-1}$, W=$-$3.9$\pm$0.4kms$^{-1}$ (Robichon et al. 1999).

Other recent extensive analyses of the velocity and age distributions of stars in the solar neighbourhood
confirm the existence of this significant substructure within the Local Association. For example, 
based on a study of nearby early-type stars, Asiain et al. (1999) resolve the supercluster into 4 
distinct components in U,V,$\tau$ space, each with a lower velocity dispersion and a more
restricted age range than the Local Association as a whole. Moving group B1, with U=$-$4.5$\pm$4.7kms$^{-1}$, 
V=$-$20.1$\pm$3.3kms$^{-1}$ and W=$-$5.5$\pm$1.9kms$^{-1}$, consists of a very young population (20$\pm
$10Myrs) and is attributable to the Scorpio-Centaurus OB association, while moving group B4, with 
U=$-$8.7$\pm$4.8kms$^{-1}$, V=$-$26.4$\pm$3.3kms$^{-1}$ and W=$-$8.5$\pm$4.7kms$^{-1}$, is substantially 
older (150$\pm$50Myrs). Given the similarities between the space velocities and ages of this moving 
group and the Pleiades open cluster, it is likely the two are related (Asiain et al. 1999).

\begin{table}
\begin{minipage}{84mm}
\begin{center}
\caption{Temperature and gravity determinations of GD50 in the refereed literature of the last 
15 years used in deriving the means. The values of Barstow et al. (1993) were excluded from the 
calculation as they lie $>$3$\sigma$ from an initial estimate of the means. We assume a minimum 
uncertainty in all temperature and gravity determinations of 2.3\% and 0.07dex, respectively.}
\label{sum1}
\begin{tabular}{ccc}
\hline
T$_{\rm eff}$(K)   & log g   & Reference \\
\hline
40538$\pm$932 & 9.22$\pm$0.08 &     1 \\
38088$\pm$876 & 8.87$\pm$0.07 &     2 \\
43102$\pm$1982 & 9.09$\pm$0.09 &    3 \\
43200$\pm$994  & 9.21$\pm$0.07 &    4 \\
39508$\pm$908  & 9.07$\pm$0.07 &    5 \\
43170$\pm$993  & 9.08$\pm$0.07 &    6 \\
41480$\pm$954  & 9.19$\pm$0.07 &    7 \\
\hline
$<$41550$\pm$720$>$ &  $<$9.15$\pm$0.05$>$ & \\
\hline

\end{tabular}
\end{center}
(1) Bergeron et al. (1991), (2) Barstow et al. (1993), (3) Bragaglia et al. (1995), (4) Vennes et al. (1997),
(5) Napiwotzki et al. (1999), (6) Aznar-Cuadrado et al. (2004) and (7) This work
\end{minipage}
\end{table}

Luhman et al. (2005) and Luhman \& Potter (2006) have recently shown that the stellar sequences of 
the AB Dor moving group (including AB Dor Ba and Bb) and the Pleiades open cluster are coincident
in M$_{\rm K}$,V$-$K and M$_{\rm K}$,J$-$K colour-magnitude diagrams. Thus not only do the AB Dor moving 
group members and the Pleiades have very similar space velocities, there is compelling evidence that the former
also have an age in the range 100-125Myrs. It is argued by these authors that the AB Dor moving group 
is thus likely to be the remnants of an OB and T association related to the formation and evolution of 
the Pleiades open cluster, a more mature version of the relationship which exists today between the 
Per OB3 open cluster and the Cassiopeia-Taurus Association (de Zeeuw et al. 1999). In this framework,
presumably the bulk of the B4 moving group is also part of this unbound remnant.
 
\subsection{GD50 and the Pleiades star formation event}

We find, in the context of the substantial velocity substructure of the Local Association, the kinematics 
of GD50 are most closely matched to those of the B4 moving group/Pleiades open cluster stream. To further
examine the possibility that GD50 is related to this particular episode of star birth we have examined it's 
location in initial mass-final mass space under the assumption that it is coeval with the Pleiades (125$\pm
$25Myrs; Ferrario et al. 2005). Noting that the determination of initial mass is extremely sensitive to the 
adopted age of the progenitor population (see Dobbie et al. 2006 for details), the proximity of GD50 to an 
extrapolation of the linear fit to the initial mass-final mass data for 27 white dwarfs in open clusters 
and the Sirius binary system (Dobbie et al. 2006, Liebert et al. 2005b, Williams, Bolte \& Koester 2005, 
Claver et al. 2001, Koester \& Reimers 1996), indicates that it's cooling age is entirely consistent with it being 
associated with this star formation event and having evolved essentially as a single star (see Figure 1). 

To get further insights we calculated the Galactic orbits of GD50 and the Pleiades back in time. We 
used the code {\sc orbit6} of Odenkirchen \& Brosche (1992) for our calculations. {\sc orbit6} computes the 
trajectory of a star in the Galactic potential for a given set of initial coordinates and space velocities.
Please refer to Pauli et al.\ (2003) for further details. For every time step we calculated the distance between
the orbits of GD50 and the Pleiades. Errors in the estimated distance and space velocities were propagated by 
means of a Monte Carlo simulation. The results of this modelling are consistent with GD50 having formed within 
the Pleiades and having been subsequently ejected. Indeed, the escape velocity from the center of the present
day cluster is only $\sim$2kms$^{-1}$ (Dehnen, priv. comm) and while the Pleiades is relatively young and one 
white dwarf member has already been firmly identified (LB1497; Luyten \& Herbig 1960), it is estimated, based 
around an extrapolation of the present day mass function, that a small number of other massive members could by
now have evolved to this configuration (e.g. Williams 2004). As dynamical evolution would have likely led the
relatively massive progenitor of this white dwarf to settle towards the most densely populated central regions of the 
cluster, it might be that, after formation, GD50 gained sufficient kinetic energy through an interaction with 
another star or a binary system to allow it to escape the Pleiades. Alternatively, some massive white dwarfs may 
receive a small recoil velocity kick during the final stages of their (super)-AGB evolution due to low level 
point-asymmetries in the outflowing material (e.g. Fellhauer et al. 2003). We note that GD50 need only have been 
moving at a mean velocity, post ejection, of $\sim$2kms$^{-1}$ with respect to the cluster for $\sim$55Myrs for
the two to be separated by $>$100pc as observed today.

A further possibility, which is less favoured but not ruled out by our Galactic orbit modelling, is that GD50 or
more particularly it's progenitor star, may have originated in an unbound OB association related to formation of 
the Pleiades open cluster. While the AB Dor moving group has only $\sim$30 known members with spectral types from 
F5-M6 (Zuckermann et al. 2004, Luhman \& Potter 2006), the B4 moving group is known to contain at least $\sim$50 
objects of spectral types A and B (Asiain et al. 1999). Furthermore, detailed simulations of the early evolution 
of a rich stellar aggregate like the nascent Pleiades, suggest that when the ignition of the OB stars drives away 
the mass dominating primordial gas, as many as 2/3 of the initial constituents may become part of an unbound and 
expanding association. Over a period comparable to the age of the B4 stream this structure can disperse over scales 
well in excess of 100pc (Kroupa, Aarseth \&  Hurley 2001).

\begin{figure}
\vspace{185pt}
\includegraphics{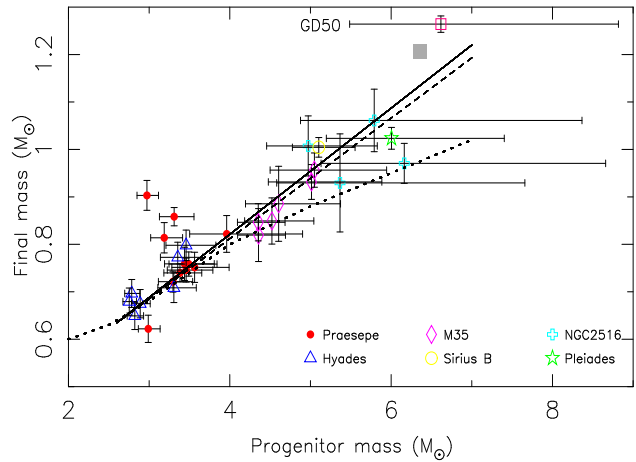}
\caption{The white dwarf members of the Hyades, Praesepe, M35, NGC2516 and Pleiades open clusters and 
the Sirius binary system in initial mass-final mass space (see text for references). A linear fit to the data 
derived by Dobbie et al. (2006) and the relation of Weidemann (2000; dotted line) are overplotted as solid 
and dotted lines, respectively. The location of GD50, derived assuming it to be coeval with the Pleiades,
is consistent with it having evolved essentially as a single star (open square). The approximate location
of PG0135+251 is also shown (filled square).}
\end{figure}

\subsection{Stellar evolution}

Although the available data do not allow us to pin down the details of the history of GD50, as a whole,
the evidence presented here provides a relatively compelling argument that it is associated with the star 
formation event that created the Pleiades. Clearly there is no need to invoke a binary white dwarf merger 
scenario to account for its evolution, if as supported by its location in Figure 1, it has evolved essentially
as a single star. Liebert et al. (2005a) have recently noted the apparent projected spatial coincidence between
many of the hot massive white dwarfs detected in the ROSAT WFC and EUVE surveys and Gould's Belt, a rich band 
of OB stars tilted at 18$^{\circ}$ with respect to the Galactic Plane which represents a ring of nearby recent 
star formation. This has led them to suggest that significantly more than 20\% of these objects may be the 
progeny of single stars. Furthermore, based on their recent determination of the form of the IFMR, Ferrario et al. 
(2005) predict that $\sim$28\% of objects in the white dwarf mass distribution have M$\simgreat$0.8M$_{\odot}$.

Indeed, the formation of ultramassive white dwarfs from single stars is predicted by the stellar evolutionary 
modelling of a number of research groups. For example, Garcia-Berro et al. (1997) show that once the helium
exhausted cores of some stars exceed $\sim$1.1M$_{\odot}$, a series of carbon-burning shell flashes can lead
to a super-AGB phase of evolution and ultimately to the formation of degenerate ONeMg cores with masses in 
the range $\sim$1.1-1.4M$_{\odot}$. Alternatively, if a star has significant rotational angular momentum, 
then the pressure lifting effect of this can allow the CO core, which will ultimately become the white dwarf, 
to grow to M$>$1.1M$_{\odot}$ (e.g. Dominguez et al. 1997). We believe the present result represents the first
compelling observational evidence directly linking ultramassive white dwarfs and single star evolution.

We also note that PG0136+251, with T$_{\rm eff}$=39640K, log g=8.99 and M=1.20$_{\odot}$ (Liebert et al. 2005a), 
has, as listed in the USNO-B1.0 catalogue, a proper motion of $\mu_{\alpha}\cos\delta$=+52$\pm$2mas yr$^{-1}$, 
$\mu_{\delta}$=$-$46$\pm$1mas yr$^{-1}$ and is thus moving on a bearing which points within 10$^{\circ}$ of the
convergent point of the Pleiades (e.g. Makarov \& Robichon 2001). Additionally, the distance calculated using
the moving clusters method (Equation 1), which assumes the white dwarf to be affiliated kinematically to this 
cluster,

\begin{equation}
d_{\rm mc}=v \sin \lambda/4.74\mu \approx 97{\rm pc}
\end{equation}

where $v$ is the space velocity of the Pleiades ($\sim$32kms$^{-1}$; Luhman et al. 2005), $\lambda$ the 
angular separation between the white dwarf and the convergent point ($\sim$95$^{\circ}$) and $\mu$ the 
tangential motion ($\sim$0.07arcsec~yr$^{-1}$), is remarkably close to the spectrophotometrically derived 
value of $\approx$93pc. The mass and the cooling time of this white dwarf can also be considered consistent 
with it having evolved essentially as a single star born at about the same time as the Pleiades open cluster 
(see Figure 1). A firm conclusion regarding the origins of PG0136+251 must, however, await a radial velocity 
measurement for this object.

Both GD50 and PG0136+251 are sometimes touted as possible examples of white dwarf binary mergers (e.g. 
Bergeron et al. 1991, Mochkovitch 1993, Segretain, Chabrier \& Mochkovitch 1997). However, the observed 
number of nearby hot ultramassive white dwarfs is difficult to reconcile with the population predicted by 
binary evolution models which adopt a plausible Galactic white dwarf merger rate (10$^{-2}$--10$^{-3}$ yr
$^{-1}$), under the assumption that these objects are the products of coalesced double degenerate systems. 
For example, for a Galactic merger rate of 10$^{-2}$yr$^{-1}$ and a cooling time scale of the order 20Myrs,
Segretain, Chabrier \& Mochkovitch (1997) estimate that the average distance to the closest hot ultramassive 
white dwarf merger product is $\sim$70pc. The probability of there being 3 within 40pc (e.g. GD50, PG1658+441; 
Green, Schmidt \& Liebert 1986, and  RE0317-854; Barstow et al. 1995) is a mere $\sim$0.1\%. However, this
difficulty must be alleviated to some extent if a 
significant proportion of these hot ultramassive white dwarfs have instead formed via single star evolution 
(Segretain, Chabrier \& Mochkovitch 1997).

A detailed study of the heliocentric space velocities of a large sample of hot young massive white 
dwarfs is clearly warranted. This would address additional questions regarding the end points of the 
evolution of intermediate mass stars, Type Ia supernovae progenitors and more generally the recent 
history of the local star formation rate.

\section[]{Summary}

We have argued using astrometric and spectroscopic data that the ultramassive white dwarf 
GD50 is likely related to the star formation event that created the Pleiades open cluster and has 
properties consistent with having evolved essentially as a single star. We believe this to be the first 
compelling observation evidence directly linking ultramassive white dwarfs and single star 
evolution. This result can help to reconcile the observed population of nearby hot ultramassive white dwarfs
with the number predicted by binary evolution models, under the assumption that these objects are the 
products of coalesced double degenerate systems.
 
\section*{Acknowledgments}
PDD and NL are sponsored by PPARC in the form of postdoctoral grants. MRB and RN 
acknowledge the support of PPARC in the form of Advanced Fellowships. We thank Walter Dehnen and Graham Wynn
for useful discussions and Nigel Hambly for forwarding to us a pre-release SuperCOSMOS proper motion 
measurement for LB1497. We are grateful to the referee for his/her input.

\appendix

\bsp

\label{lastpage}

\end{document}